\documentclass[9pt,twocolumn, twoside]{osajnl}

\journal{ol} % Choose journal (ao, aop, josaa, josab, ol, pr)

\setboolean{shortarticle}{true}                                                              

\usepackage[super]{nth}
\usepackage{setspace}
\title{\vspace{-2cm} Reciprocal plasmonic metasurfaces: Theory and applications}

\author[1,*]{Yanzeng Li}
\author[1]{Micheal McLamb}
\author[1]{Serang Park}
\author[2]{Darrell Childers}
\author[1]{Glenn D.~Boreman}
\author[1,3]{Tino Hofmann}

\affil[1]{Department of Physics and Optical Science, University of North Carolina at Charlotte, 9210 Univ. City Blvd., Charlotte, NC 28223, USA}
\affil[2]{USCONEC,1138 25th Street Southeast, Hickory, NC 28602, USA}
\affil[3]{THz Materials Analysis Center, Department of Physics, Chemistry, and Biology (IFM), Link{\"o}ping University, SE 581 83 Link{\"o}ping, Sweden}

\affil[*]{Corresponding author: yli91@uncc.edu}

\begin{abstract}
A new configuration for metasurface construction is presented to achieve multi-functional capabilities including perfect absorption, bio/chem sensing, and surface-mode lasing. The reciprocal plasmonic metasurfaces discussed here are composed of two plasmonic surfaces of reciprocal geometries separated by a dielectric spacer. Compared to conventional metasurfaces this simple geometry exhibits an enhanced optical performance. The discussed reciprocal metasurface design further enables effective structural optimization and allows for a simple and scalable fabrication. The physical principle and potential applications of the reciprocal plasmonic metasurfaces are demonstrated using numerical and analytical approaches.
\end{abstract}

\setboolean{displaycopyright}{true}

\begin{document}

\maketitle

%%%%%%%%%%%%%%%%%%%%%%%%%%%% Introduction %%%%%%%%%%%%%%%%%%%%%%%%%%%%%%%
Metasurfaces are resonant textured surfaces with unit cells much smaller than the wavelength of incident electromagnetic radiation \cite{Shalaev2009}. One of the most frequently used metasurface designs are heterostructures composed of triple-layered constituents which are optimized for optical frequencies \cite{landy2008perfect,chen2010antireflection,ameling2010cavity,shchegolkov2010perfect,pu2011design,vazquez2014hybrid}. Such heterostructured metasurfaces have been extensively studied over the past decade and exhibit excellent optical performance \cite{ye2010omnidirectional,wu2011large,zhang2011polarization,chen2015infrared,feng2012engineering,shrestha2018indium,liu2010infrared,artar2009fabry,lu2015metal}.

Despite these advances, the explored designs often require complex nanofabrication processes which can lead to a reduced performance of the as-synthesized metasurfaces compared to numerical results obtained from optimized, nominal designs \cite{dao2019selective}. 
For example, one of the most common fabrication defects when using a mask or lift-off processes is the dislocation of elements of the metasurface \cite{dao2015infrared, dao2019selective}. Furthermore, loss in fidelity often results in an imperfect reproduction of the nominal geometries of the metasurfaces constituents \cite{dao2015infrared, dao2019selective, d2014large}. These imperfections directly impact the performance of the heterostructured metasurfaces. For instance, in metasurfaces designed as perfect absorbers, the experimentally observed absorption values are often much lower \cite{dao2015infrared}. In addition, fabrication-induced imperfections can lead to grain boundaries between areas with different defect densities, which can further diminish the performance of the entire metasurface \cite{dao2019selective}. 

% Now what are we doing to solve this
Here we introduce a metasurface design that can be synthesized using a very simple two-step fabrication process. The process consists of the deposition of a structured polymer layer and a subsequent metallization. This results in two metal-based metasurfaces with reciprocal geometry separated by a dielectric spacer.
%Here simple dipoles structures are investigated. 

%What are we doing and what are the results?
The reciprocal plasmonic metasurface is theoretically studied here using numerical and analytical effective medium optical models. It is found that reciprocal plasmonic metasurface designs exhibit unique advantages for applications in perfect absorption, sensing of minute index changes, and surface-mode light emission, compared to conventional heterostructured metasurfaces. The record values obtained for the figure of merit for index sensing suggest that these structures are ideal candidates for extremely sensitive ambient index sensors \cite{ameling2010cavity}. In addition to providing enhanced optical performance, the reciprocal plasmonic metasurface discussed here dramatically reduces the fabrication complexity compared to the currently existing heterostructured metamaterials.

\begin{figure}[t]
	\centering
	\includegraphics[width=0.9\linewidth]{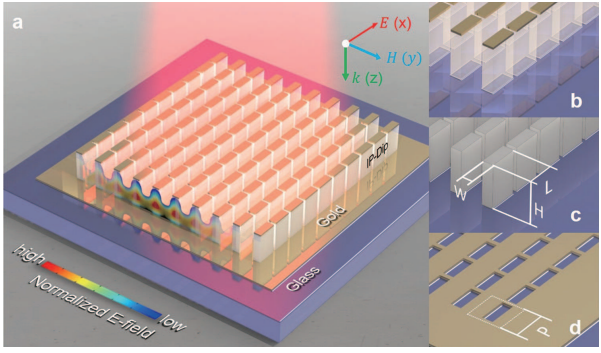}
	\caption{(a) The reciprocal metasurface composed of a three-layered heterostructure: Au bar-antenna array (b), polymer-based fin-array (c), and a patterned Au surface reciprocal to the rectangular bar array (d). \vspace{-.3cm} }
	\label{fig:NominalDesign}
\end{figure}

%%%%%%%%%%%%%%%%%%%%%%%%%%%% Main context %%%%%%%%%%%%%%%%%%%%%%%%%%%%%%%
The geometry of this reciprocal metasurface is depicted in  Fig.~\ref{fig:NominalDesign}. The top plasmonic metasurface is composed of rectangular bars while the bottom metasurface has a reciprocal arrangement with rectangular openings in a metallized surface. The dielectric spacer is formed by fins with a rectangular base.

\begin{figure}[t]
	\centering
	\includegraphics[width=0.99\linewidth, keepaspectratio=true, trim=0 0 0 0, clip]{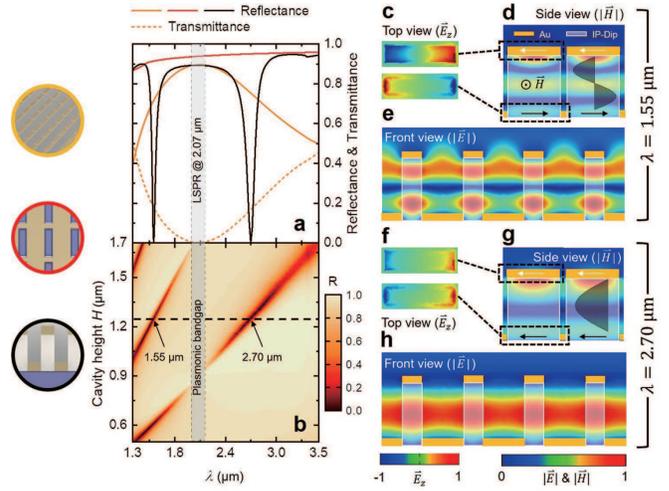}
	\caption{(a) Numerically calculated reflectance (solid) and transmittance (dotted) of the optimized reciprocal plasmonic metasurface (black) and its individual constituent parts: Au bar-antenna array (orange) and perforated film (red). (b) Numerically determined reflectance map of the reciprocal metasurface as a function of the height $H$ of the fins valued from 0.5~$\mu$m to 1.7~$\mu$m. Panel (c) and (d) display the normalized electric field distributions $|\vec{E}|$ at the resonant wavelengths of 1.55~$\mu$m and 2.70~$\mu$m, respectively, as indicated by the black dashed line in panel~(b). Panels (c) - (e) and (f) - (h) depict a second-order harmonic and fundamental harmonic standing wave for the wavelengths of 1.55~$\mu$m and 2.70~$\mu$m, respectively.  \vspace{-.3cm}  }
	\label{fig:NumOptMod}
\end{figure}

\begin{figure}[ht]
	\centering
	\includegraphics[width=0.9\linewidth, keepaspectratio=true, trim=0 40 0 0, clip]{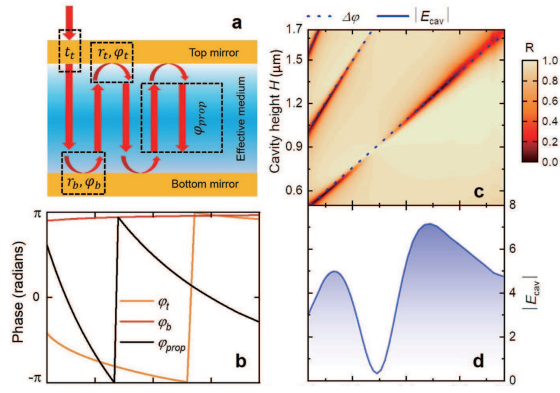}
	\caption{(a) Effective medium optical model of the optimized reciprocal metasurface consists of an effective medium sandwiched between two effective mirrors. Panel (b) shows phase changes at the mirrors ($\varphi_{\text{t}}$,$\varphi_{\text{b}}$) and due to the propagation ($\varphi_{\text{prop}}$). (c) Calculated resonant wavelengths (black dashed lines) with varying quantities of the parameter $H$ from 0.5~$\mu$m to 1.7~$\mu$m, wherein the numerically determined reflectance map shown in Fig.~\ref{fig:NumOptMod}~(b) is shown for comparison. Panel (d) shows the calculated relative amplitude of the confined electric field $|E_{\text{cav}}|$ inside the effective medium.  \vspace{-.3cm} }
	\label{fig:EffOptMod}
\end{figure}

The optical performance of the reciprocal metasurface is optimized by carefully engineering the geometrical parameters using a numerical optical model based on the finite-element method (FEM) as shown in Fig~\ref{fig:NumOptMod}. The reciprocal metasurface was developed to achieve a narrow and strong absorption at a wavelength of 1.55~$\mu$m and a quality factor of 20, which is relevant for telecommunication applications. In addition, the physical mechanism of the reciprocal metasurface has been theoretically studied by an effective  medium optical model based on interference theory \cite{chen2012interference}, as shown in Fig.~\ref{fig:EffOptMod}.

The reciprocal metasurface designed here consists of a dielectric fin-array sandwiched by two Au patterned layers, as shown in Fig.~\ref{fig:NominalDesign}. 
This array is composed of fins with the following  
dimensions: length $L=0.9~\mu$m, width $W=0.3~\mu$m, periodicity $P=1~\mu$m, and height $H=1.25~\mu$m in a square lattice pattern, as shown in Fig.~\ref{fig:NominalDesign} (c).
This design allows rapid protoyping using two-photon polymerization and while the simplicity of the geometry enables the fabrication of larger surface areas using standard lithography techniques \cite{Y.Li2018BroadbandAntireflection}. 
For the calculations discussed below, it was assumed that the top bar resonator array and bottom perforated film are fabricated from gold with a thickness of 50~nm. The dielectric spacer is assumed to be composed of IP-Dip, a polymer compatible with the two-photon polymerization process, for which the optical properties are well known in the infrared and visible spectral range \cite{li2019uv}.

Fig.~\ref{fig:NumOptMod} (a) shows the calculated reflectance spectrum (black solid line) of the optimized reciprocal metasurface under normal incidence illumination with the electric field polarized along the long axis of the bars (x-axis). In the spectral range from 1.3~$\mu$m to 3.5~$\mu$m, two nearly-zero-valued minima centered at 1.55~$\mu$m and 2.70~$\mu$m can be observed.   

For comparison, the response of the top metasurface (rectangular bar array, orange solid line) and the bottom metasurface (rectangular hole array, red solid line) are also depicted in Fig.~\ref{fig:NumOptMod} (a).   

The normalized electric and magnetic field distributions $|\vec{E}|$ and $|\vec{H}|$, respectively, 
corresponding to the two resonant wavelengths are shown in Fig.~\ref{fig:NumOptMod}~(c) - (h), respectively. 
The field distributions for $H=1.25~\mu$m in Fig.~\ref{fig:NumOptMod}~(c) - (e) and (f) - (h) depict a second-order harmonic and fundamental harmonic standing wave for the wavelengths of 1.55~$\mu$m and 2.70~$\mu$m, respectively.
The locations of the reflectance minima depend linearly on the the cavity height $H$. This is illustrated in the reflectance map shown in Fig.~\ref{fig:NumOptMod}~(b) where the reflectance is plotted as a function of $H$ valued from $0.5~\mu$m to $1.7~\mu$m over the spectral range from $1.3~\mu$m to $3.5~\mu$m. Fig.~\ref{fig:NumOptMod}~(b) also indicates an additional higher-order mode, which can be accessed if the dielectric fin-array has a sufficient height. For the investigated spectral range, higher-order modes can be observed for $H>0.9\mu$m. It can be concluded that with increasing $H$, the resonant modes of the reciprocal metasurface red-shift while higher-order modes start to appear at the short wavelength end of the spectrum. 

In addition to being a function of the cavity height $H$, the optical response of the reciprocal metasurface also depends on the plasmon resonance of the Au bar-antenna array. The reflectance (orange solid line) and transmittance (orange dotted line) spectra of the top Au bar-antenna array are included in Fig.~\ref{fig:NumOptMod}~(a) for comparison indicating a plasmon resonance at 2.07~$\mu$m. Note, that the modes of the reciprocal metasurface are interrupted in the vicinity of the plasmon resonance as can be clearly seen in the reflectance map [Fig.~\ref{fig:NumOptMod}~(b)]. This is due to the lack of transparency of the top metasurface as shown in Fig.~\ref{fig:NumOptMod}~(a) which prevents the effective coupling to the bottom metasurface. 

The reflectance of perforated Au film at the bottom of the reciprocal metasurface [red solid line, Fig.~\ref{fig:NumOptMod} (a)], on the other hand, shows very little variation as a function of the wavelength. This observation is relevant for the development of an analytical model discussed below where the perforated Au film at the bottom of the reciprocal metasurfaces is treated as a plasmonic mirror reflecting all the wavelengths of interest.

shown in Fig.~\ref{fig:NumOptMod} (a) defines the optical property of the perforated Au film individual of the other layers (red). Upon this observation, it is appropriate to treat the perforated Au film as a plasmonic mirror reflecting all the wavelengths within the band of interest.

Based on the results of the FEM-based calculations, an analytical effective medium-based formalism is developed in the following. In contrast to the FEM-based numerical model the analytical model allows the rapid determination of the geometry parameters of the reciprocal metasurface which are required to obtain specific target frequencies for its absorption bands. In addition, this model further provides insights into the physical mechanism which is driving the optical response of the reciprocal metasurfaces discussed here.

%needs more work, just rough
Figure~\ref{fig:EffOptMod}~(a) depicts the concept of the effective medium optical model. Due to the sub-wavelength-sized structural features of the dielectric fin-array [as shown in Fig.~\ref{fig:NominalDesign} (c)], it can be treated as a single layer with an effective index $n_{\text{eff}}$. $n_{\text{eff}}$ can be calculated using the Bruggeman effective medium approximation and the complex dielectric function of the constituent material IP-Dip \cite{Y.Li2018BroadbandAntireflection,li2018high,Fullager2017}. This layer is enclosed by two reflective surfaces (top Au bar array and bottom Au perforated film) which are denoted in Fig.~\ref{fig:EffOptMod}~(a) as top and bottom mirror. 

As the incident electromagnetic radiation transmits through the top partial-transparent mirror [Fig.~\ref{fig:EffOptMod} (a)], a cavity mode forms inside the effective medium layer. This phenomenon only occurs when these waves oscillate in phase, i.e., the phase difference $\Delta\varphi$ after a round trip are integer multiples of 2$\pi$:

\begin{equation}    \label{eqn:1}
\Delta\varphi=\varphi_{\text{t}}+\varphi_{\text{b}}+\varphi_{\text{prop}}=N2\pi~~~~(N=1, 2,...),
\end{equation}

\noindent where $\varphi_{\text{t}}$ and $\varphi_{\text{b}}$ respectively denote the phase changes induced by the interaction of the electrical radiation with the top and bottom mirrors, for which the numerical calculations on each individual part determine their spectra shown in Fig.~\ref{fig:EffOptMod}~(b). The phase change due to the propagation in the effective medium $\varphi_{\text{prop}}$ is calculated by $\varphi_{\text{prop}}=2\pi Hn_{\text{eff}}/\lambda$ for $H=1.25~\mu$m and also included in Fig.~\ref{fig:EffOptMod}~(b) (black solid line).

Varying the value of the parameter $H$ in the same range as the one used in the numerical simulation shown in Fig.~\ref{fig:NumOptMod}~(b), Eqn.~(\ref{eqn:1}) allows the identification of the cavity resonances with $N=1,$ 2, and 3 which are depicted by the black dotted lines in Fig.~\ref{fig:EffOptMod}~(c).
Note, that the results for the reflectance map of the FEM-based model [Fig.~\ref{fig:NumOptMod}~(b)] are reproduced here again for direct comparison. 

A good agreement between the analytically calculated and the numerically simulated results is achieved for the location of the reciprocal metasurface modes. This confirms the optical-cavity-based mechanism of the reciprocal metasurface.

\begin{figure}[tb]
	\centering
	\includegraphics[width=\linewidth]{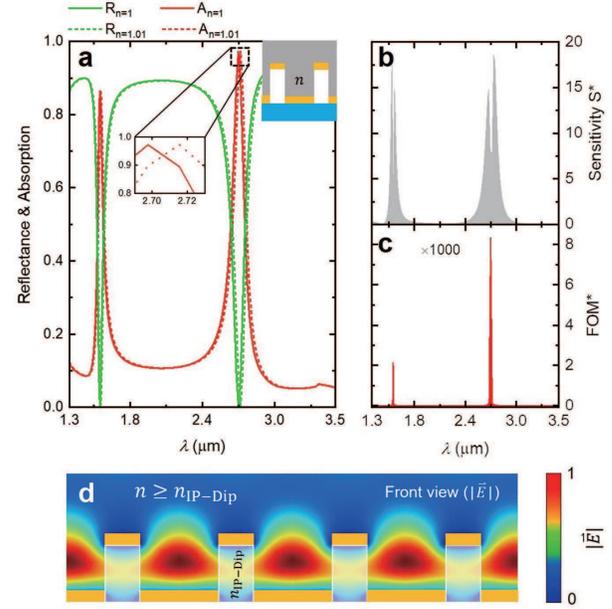}
	\caption{(a) Nearly perfect absorption with a Q-factor of 20 is obtained at the resonant wavelength of 2.70~$\mu$m (red solid line). As ambient index $n$, as shown in the inset, deviates from 1 to 1.01, the resonance peak at 2.70~$\mu$m has a 0.02~$\mu$m red shift (red dashed line), shown in the zoomed-in plot. The sensitivity S$^*$ and figure of merit FOM$^*$ are shown in (b) and (c) for characterizing the sensing capability of the reciprocal plasmonic metasurface. Panel (d) indicates the normalized electric field $|\vec{E}|$ when $n\geq n_{\text{IP-Dip}}$.  \vspace{-.3cm} }	
	\label{fig:Appl}
\end{figure}

The reflectance map [Figs.~\ref{fig:NumOptMod}~(b) and \ref{fig:EffOptMod}~(c)] further revealed that the strength of the cavity modes are wavelength dependent, an effect which can not be obtained using a simple phase analysis as shown in Eqn.~(\ref{eqn:1}).
Instead, this effect can be described considering the relative amplitude $|E_{\text{cav}}|$ of an infinite number of electric fields that constructively interfere inside the layer with the effective index $n_{\text{eff}}$:

\begin{equation}    \label{eqn:2}
\left|E_{\text{cav}}\right|\propto\frac{t_{\text{t}}^2}{1+\mathopen|r_{\text{t}}^2r_{\text{b}}^2\mathclose|},
\end{equation}

\noindent where $t_{\text{t}}$,$r_{\text{t}}$ and $r_{\text{b}}$ denote the transmissivity and reflectivity for the top and bottom mirrors, respectively, and their values are derived from the corresponding transmittance and reflectance spectra obtained by the FEM-based calculations [Fig.~\ref{fig:NumOptMod}~(a)]. 

$|E_{\text{cav}}|$ is shown in Fig.~\ref{fig:EffOptMod}~(d) for the spectral range from 1.3 to 3.5~$\mu$m. Two maxima centered at $1.60~\mu$m and $2.66~\mu$m can be observed. The minimum at $2.07~\mu$m corresponds to the plasmon resonance of the top mirror.  
Therefore, the pair of the plasmonic surfaces function in a way that not only alters the phase change at the interface but also modulates the amplitude of the electromagnetic radiation trapped inside the cavity, consequently being able to influence both resonance position and resonant strength.

So far, a comprehensive understanding of the optical mechanism of the reciprocal metasurface has been achieved. In the following potential applications are briefly explored. 

Fig.~\ref{fig:Appl}~(a) depicts the absorption spectrum (red solid line) of the optimized reciprocal metasurface. Two absorption peaks at wavelength of 1.55~$\mu$m and 2.70~$\mu$m can be observed. The Q-factor for both absorption peaks is approximately 20. Therefore the reciprocal metasurface could be readily applied as a narrow-band perfect absorber in the near-infrared spectral range for which the absorption frequencies can be easily adjusted by the geometry and the refractive index of the dielectric spacer of the reciprocal metasurface. 

Furthermore, this type of metasurface is an excellent candidate for sensing applications due to its high sensitivity to the ambient refractive index $n$, as shown in the inset of Fig.~\ref{fig:Appl}~(a). An index change of 0.01 would result in a shift of the resonant wavelength of 0.02~$\mu$m which can be easily detected.
Its sensing capability is characterized in terms of sensitivity S$^{*}$, defined as the ratio of the intensity change vs. the change in refractive index, while the figure of merit FOM$^{*}$ is the ratio of S$^{*}$ vs. the absolute intensity \cite{ameling2010cavity}. For the example depicted in Fig.~\ref{fig:Appl}~(b) and (c) S$^{*}$ is a approximately 18, while FOM$^{*}$ is found to be larger than 8000. These values are substantially higher than those obtained for other metasurface geometries and in fact would set a new record for heterostructured surface sensors \cite{ameling2010cavity}. This extreme sensitivity is due to the strong change of the plasmonic resonance of the top and bottom metasurfaces upon ambient index variation. Furthermore, the effective refractive index of the cavity $n_{\text{eff}}$ also varies as a function of the ambient refractive index, which amplifies the changes in the optical response of the reciprocal metasurface.  

In addition to the discussed applications for sensing and as a perfect absorber, the results of the FEM-based optical model simulation imply that the reciprocal metasurface could be used for surface mode lasing.
For this application, a gain medium is integrated into the reciprocal metasurface, as shown in Fig.~\ref{fig:Appl}~(d). Assuming the space between the fins is filled with a gain material with the index $n$ larger than or comparable to that of the polymer fins $n_{\text{IP-Dip}}$, a strong field enhancement, indicated by the normalized electric field $|\vec{E}|$, can be observed between the fins. This enhancement effect can be exploited for surface-mode lasing in conjunction with the partial transparency of the Au bar-antenna array, which serves as the top mirror of the laser cavity.

%%%%%%%%%%%%%%%%%%%%%%%%%%%% Conclusion %%%%%%%%%%%%%%%%%%%%%%%%%%%%%%%

To conclude, the optical properties of a reciprocal plasmonic metasurface were theoretically studied using a FEM-based approach and an effective medium optical model. The near-infrared optical response of the reciprocal metasurface discussed here reveals strong resonances for which almost perfect absorption is found, assuming realistic dielectric properties of the constituents. It is demonstrated that the resonance frequencies can be easily tuned by changing the height of the dielectric spacer between the top bottom Au layers with reciprocal geometries.

The results of the FEM-based calculations are in very good agreement with the data obtained from the analytical, effective medium based model. In comparison with the FEM-based model, the analytical effective-medium-based optical model allows the rapid optimization of the geometry of the reciprocal metasurface for a wide range of target frequencies. The analytical model further reveals the cause of the strong absorption peaks, which is an interplay between an effective optical-cavity and the plasmon resonances of the reciprocal top and bottom surfaces. 

The theoretical findings presented here illustrate the advantages of the reciprocal plasmonic metasurfaces in the applications of perfect absorption. The narrow absorption characteristic and its strong dependence on the ambient refractive index further suggest that reciprocal metasurfaces can be employed as optical sensors. Comparing the sensitivity with other metasurfaces used for sensing applications showed that record high figures of merit may be achieved using the reciprocal metasurface design. 

In addition to applications for perfect absorption and sensing of minute ambient refractive index changes, normalized electric field calculations exhibit an enhancement effect which could be exploited for the fabrication of surface-mode lasers. This application would require the use of a suitable gain medium with a higher refractive index than the dielectric spacer of the reciprocal metasurface. 

While studied theoretically here, reciprocal metasurfaces have substantial practical implications as the simplicity of its design suggests a considerable reduction in fabrication complexity often plaguing heterostructured metasurfaces demonstrated so far. In fact, the discussed geometry is suitable for a simple two step fabrication. In the first step the dielectric spacer would be deposited, for instance using mask-less two-photon polymerization approaches. In the subsequent metallization step the reciprocal top and bottom surfaces can be deposited at the same time. 

We believe that the reciprocal plasmonic metasurfaces can open a new avenue for the design and fabrication of metasurfaces with enhanced performance yet low manufacturing time and cost.

%%%%%%%%%%%%%%%%%%%%%%%%%%%% Funding %%%%%%%%%%%%%%%%%%%%%%%%%%%%%%%
\vspace{-0.1cm}
\section*{Funding Information}
The authors are grateful for support from the National Science Foundation (1624572) within the IUCRC Center for Metamaterials and through the NSF MRI 1828430, the Army Research Office (W911NF-14-1-0299) and the Department of Physics and Optical Science of the University of North Carolina at Charlotte.

\vspace{-0.1cm}
\section*{Disclosures}
The authors declare no conflicts of interest.

\vspace{-0.1cm}
% Bibliography
%\bibliography{CompleteLibrary_Li}

% Full bibliography will be added automatically on a new page for Optics Letters submissions. This command is ignored for journal article submissions.
% Note that this extra page will not count against page length.
%\bibliographyfullrefs{CompleteLibrary_Li}

%Manual citation list
%\begin{thebibliography}{1}

%\bibitem{Zhang:14}
%Y.~Zhang, S.~Qiao, L.~Sun, Q.~W. Shi, W.~Huang, %L.~Li, and Z.~Yang,
% \enquote{Photoinduced active terahertz metamaterials with nanostructured
%vanadium dioxide film deposited by sol-gel method,} Opt. Express \textbf{22},
%11070--11078 (2014).
%\end{thebibliography}

\end{document}